\documentclass[prl, twocolumn, showpacs, preprintnumbers, amsmath, amssymb] {revtex4-1}
\usepackage{graphicx} 
\usepackage{dcolumn} 
\usepackage{mathrsfs}
\usepackage{bm} 

\begin{document}

\preprint{}

\title{Quantum-State Purity of Heralded Single Photons Produced from Frequency-Anti-Correlated Biphotons}

\author{Shengwang Du}\email{Corresponding author: dusw@ust.hk}
\affiliation{Department of Physics, The Hong Kong University of Science and Technology, Clear Water Bay, Kowloon, Hong Kong, China}

\date{\today}

\begin{abstract}
{We analyze the quantum-state purity of heralded single photons produced from frequency-anti-correlated biphotons. We find that the quantum-state purity in time-frequency domain depends strongly on the response time uncertainty of the trigger-photon detector that heralds the generation of its paired photon. If the trigger response time is much shorter than the two-photon coherence time, the time-frequency quantum-state purity of heralded single photons approaches unity and the heralded single photon is in a nearly pure state. If the trigger response time is much longer than the two-photon coherence time, the heralded photon is then projected onto a mixed state. Making use of the time-frequency entanglement, heralded single photons with a well-defined temporal wave function  or a frequency superposition state can be produced and engineered. This time-frequency entanglement allows for shaping heralded single photons through nonlocal spectral modulation.}
\end{abstract}





\maketitle


Correlated photon pairs can be used to generate heralded single photons: the detection of one photon heralds the presence of the remaining one and projects it onto a single-photon Fock state. Spontaneous parametric down conversion (SPDC, $\chi^{(2)}$ nonlinear process) \cite{HarrisPRL1967, WeinbergPRL1970} and spontaneous four-wave mixing (SFWM, $\chi^{(3)}$ nonlinear process) \cite{SFWM01, DuJOSAB2008} have been two standard methods for producing paired photons. As driven by continuous-wave (CW) pump laser fields, the photon pairs generated from these parametric processes are time-frequency entangled because of the energy conservation raised from  the time-translation symmetry, \textit{i.e.,} the sum of the frequencies of paired two photons is fixed while individual photons have their bandwidths. It has been commonly believed that, this frequency anti-correlation would project the heralded single photon into a mixture of frequency modes in its bandwidth \cite{GricePRA2001, MosleyPRL2008}, and spectral filtering to obtain an approximate pure single-photon state results in a dramatically reduced photon rate by discarding photons outside of the filter frequency mode \cite{ZeilingerPRL2006, PanNP2007}.  A complete different approach to obtain a heralded pure single-photon state is thus eliminating the entanglement and generating photon pairs with factorable (frequency-uncorrelated) joint spectrum driven by pulsed lasers \cite{GricePRA2001, URenLP2005, MosleyPRL2008, MigdallOE2010, KimPRL2014, DLCZ, LukinScience2003, KimbleNature2003, KimbleOE2006}.

For frequency-anti-correlated biphotons, the heralding detection is commonly modeled  as tracing over the trigger photon and thus reduces the heralded photon into a mixed state \cite{URenLP2005}. In this Letter, we point out that an ideal heralding process with an instantaneous trigger-photon detection projects the remaining frequency-anti-correlated photon onto a pure quantum state, which is a superposition state of its frequency components. When the trigger photon detection has a finite response time, it degrades the purity of the single photons. We analyze the quantum-state purity of heralded single photons produced from frequency-anti-correlated biphotons. If the trigger response time is much longer than the two-photon coherence time, the heralded photon is projected onto a mixed state, as well known for decades. If the trigger response time is much shorter than the two-photon coherence time, we find that, the time-frequency quantum-state purity of heralded single photons approaches nearly unity without any need of spectral filtering.

Let's start with the biphoton state (in time-frequency space) of a frequency anti-correlated photon pair
\begin{eqnarray}
\label{eq:BiphotonState}
|\Psi_{12}\rangle=\int d\Omega \Phi(\Omega)\hat{a}^\dagger_{2}(\omega_{20}-\Omega)\hat{a}^\dagger_{1}(\omega_{10}+\Omega)|0\rangle,
\end{eqnarray}
where $|0\rangle$ is the vacuum state, $\omega_{10}$ and $\omega_{20}$ are the central angular frequencies of photons 1 and 2. $\hat{a}^\dagger_{1}$ and $\hat{a}^\dagger_{2}$ are the field creation operators. $\Phi(\Omega)$ is the two-photon joined spectrum function. The frequency entanglement of the biphoton state is a result of the energy conservation $\omega_{1}+\omega_{2}=\omega_{10}+\omega_{20}$. The field operators in time domain can be expressed as:
\begin{eqnarray}
\label{eq:FieldOperator_t}
\hat{a}_i(t)=\frac{1}{\sqrt{2\pi}}\int d\omega \hat{a}_i(\omega)e^{-i\omega t}.
\end{eqnarray}
The field operators satisfy the commutation relations $[\hat{a}_i(\omega), \hat{a}^\dagger_{j}(\omega')]=\delta_{ij} \delta(\omega-\omega')$ and $[\hat{a}_i(t), \hat{a}^\dagger_{j}(t')]=\delta_{ij}\delta(t-t')$.  In the state described in Eq. (\ref{eq:BiphotonState}), the two photons are perfectly paired. It can be shown that the photon pair generation rate $R$ and the individual single-photon rates $R_i$ are equal and time invariant:
\begin{eqnarray}
\label{eq:Rate}
R=R_i=\langle\Psi_{12}|\hat{a}^\dagger_{i}(t)\hat{a}_{i}(t)|\Psi_{12}\rangle=\frac{1}{2\pi}\int d\Omega |\Phi(\Omega)|^2,
\end{eqnarray}
because the system is driven by CW pump fields and has the time-translation symmetry.

Detection of the photon 2 at time $t_2$ reduces the two-photon state in Eq. (\ref{eq:BiphotonState}) to the folowing state:
\begin{eqnarray}
\label{eq:HeraldedState0}
|\Psi_{1}\rangle_2=\frac{1}{\sqrt{R}}\hat{a}_{2}(t_2)|\Psi_{12}\rangle,
\end{eqnarray}
which is the heralded single-photon state with the normalization factor $1/\sqrt{R}$. To prove this is a pure quantum state, we look at its density operator
\begin{eqnarray}
\label{eq:DensityOperator0}
\hat{\rho}_{1|2}=|\Psi_{1}\rangle_{22} \langle \Psi_{1}|.
\end{eqnarray}
It is obvious that
\begin{eqnarray}
\label{eq:DensitySquare0}
\hat{\rho}^2_{1|2}=\hat{\rho}_{1|2},
\end{eqnarray}
which shows the heralded single-photon state in Eq. (\ref{eq:HeraldedState0}) is indeed a pure quantum state. Therefore, we have proved that in the ideal heralding process described above, where the detection of the trigger photon 2 has an infinitely short response time and there is no time uncertainty in the heralding process, the projected single photon is in a pure state.

Now we turn to the real situation where the the trigger-photon detector has a finite response time $\Delta t$: one does not know the exact time origin of the heralded single photon during this response time window. In this case, the density operator can be expressed as
\begin{eqnarray}
\label{eq:DensityOperator}
\hat{\bar{\rho}}_{1|2}&=&\frac{1}{\Delta t}\int_{-\Delta t/2}^{\Delta t/2} dt_2 |\Psi_{1}\rangle_{22} \langle \Psi_{1}| \\ \nonumber
&=&\frac{1}{R \Delta t}\int_{-\Delta t/2}^{\Delta t/2} dt_2 \hat{a}_{2}(t_2)|\Psi_{12}\rangle \langle \Psi_{12}|\hat{a}^\dagger_{2}(t_2).
\end{eqnarray}
Making use of Eq. (\ref{eq:FieldOperator_t}), the density operator can be rewritten as
\begin{eqnarray}
\label{eq:DensityOperator2}
\hat{\bar{\rho}}_{1|2}=\frac{1}{2\pi R}\int d\omega_2 d\omega_2' \textrm{sinc}\big[\frac{(\omega_2-\omega_2')\Delta t}{2}\big] \\ \nonumber
\times \hat{a}_{2}(\omega_2)|\Psi_{12}\rangle \langle \Psi_{12}|\hat{a}^\dagger_{2}(\omega_2').
\end{eqnarray}
We then get the density matrix element in the frequency domain
\begin{eqnarray}
\label{eq:DensityMatrixElement}
\bar{\rho}_{1|2}(\omega_1, \omega_1')=\langle0|\hat{a}_1(\omega_1) \hat{\bar{\rho}}_{1|2} \hat{a}^\dagger_1(\omega_1')|0\rangle \nonumber \\
=\frac{1}{2\pi R}\textrm{sinc}\big[\frac{(\omega_1-\omega_1')\Delta t}{2}\big] \Phi(\omega_1-\omega_{10}) \Phi^*(\omega_1'-\omega_{10}).
\end{eqnarray}
The quantum-state purity of the heralded single photons can be computed from $\gamma=\textrm{Tr}(\hat{\bar{\rho}}^2_{1|2})$ \cite{URenLP2005, YinPRA2008}. With the frequency-entangled biphoton state in Eq. (\ref{eq:BiphotonState}), we obtain the purity of its heralded single-photon state
\begin{eqnarray}
\label{eq:Purity}
\gamma=\frac{1}{(2\pi R)^2}\int d\Omega d\Omega'\textrm{sinc}^2\big[\frac{(\Omega-\Omega')\Delta t}{2}\big] \nonumber \\
\times|\Phi(\Omega)\Phi(\Omega')|^2
\end{eqnarray}

Obviously, as the trigger-photon detection is instantaneously fast ($\Delta t\rightarrow 0$), the density operator is reduced to that of a pure state $\hat{\bar{\rho}}_{1|2}\rightarrow\hat{\rho}_{1|2}(t_2=0)$ and we have an unity purity ($\gamma=1$). This is consistent with our previously described ideal heralding process that projects the remaining photon onto a pure quantum state. As the trigger-photon detector is ultraly slow ($\Delta t\rightarrow \infty$), $\textrm{sinc}\big[\frac{(\omega_1-\omega_1')\Delta t}{2}\big] \rightarrow 2\pi\delta(\omega_1-\omega_1')/\Delta t$, all non-diagonal density matrix elements in Eq. (\ref{eq:DensityMatrixElement}) vanishes and the heralded single photon is in a completely mixed state.

For the case with a finite $\Delta t$, the quantum-state purity is characterized by Eq. (\ref{eq:Purity}). As the trigger-photon detection response time is much shorter than the two-photon coherence time (inverse of the bandwidth of the joint spectrum) such that $(\Omega-\Omega')\Delta t \ll 1$ holds within the bandwidth, we have $\gamma \sim 1$ and the heralded photon is in a nearly pure state. To show how the purity of the heralded single photon state is affected by the finite $\Delta t$, we consider two examples in the following.


\begin{figure}
\includegraphics[width=\linewidth]{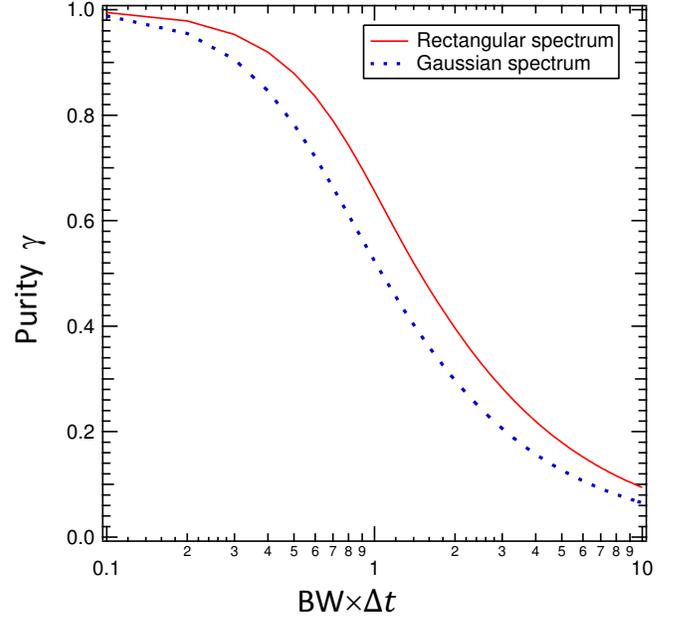}
\caption{(color online). Quantum-state purity $\gamma$ as a function of BW$\times \Delta t$, the product of biphoton bandwidth (BW) and the trigger-photon detection response time ($\Delta t$). The solid (red) curve is calculated from the rectangular-shape spectrum, and the dotted (blue) curve from the Gaussian-shape spectrum with BW as the full width at half maximum.
\label{fig:purity}}
\end{figure}

\textit{Case 1 with a rectangular-shape spectrum.}  In the first, we consider transform-limited biphotons with maximum frequency entanglement within their angular-frequency bandwidth of $2\pi BW$. In this case, the joint spectrum function $\Phi(\Omega)=\sqrt{R/BW}$ is nonzero only within the bandwidth $\Omega\in[-\pi BW, \pi BW]$. The solid curve in Fig. \ref{fig:purity} shows the numerical result of the purity as a function of the trigger response time. At $BW \Delta t \leq 0.1$, the purity $\gamma \geq 0.99$. The purity decreases as we increase the response time, but not very badly. As the response time approaches the two-photon coherence time ($BW \Delta t = 1$), the purity is still as high as 0.66.  As we further increase the response time to make $BW \Delta t = 10$,  the purity drops down to 0.09.

\textit{Case 2 with a Gaussian-shape spectrum.}  In this case, the joint spectrum function can be expressed as $\Phi(\Omega)=\sqrt{\frac{4R\sqrt{\pi \ln 2}}{2\pi BW}}e^{-2\ln 2 (\Omega/2\pi BW)^2}$, where $BW$ is the full width at half maximum of $|\Phi(\Omega)|^2$. The numerical result of the purity as a function of the trigger response time is plotted as the dotted curve in Fig. \ref{fig:purity}, which is comparable to that of Case 1 with a rectangular-shape spectrum. At $BW \Delta t = 0.1$, the purity $\gamma = 0.99$. At $BW \Delta t = 1$, the purity becomes 0.52. The purity drops down to 0.07 as we further increase the response time to make $BW \Delta t = 10$.

In both cases, the quantum-state purity $\gamma > 0.98$ at $BW \Delta t=0.1$, and holds well above 0.90 for $BW \Delta t< 0.3$. We can treat such heralded photon as a pure single-photon state. For biphotons generated from SPDC or SFWM in solid-state materials without spectral filtering or cavity enhancement, their bandwidths are normally much wider than THz. As a result, it is impossible to heralding pure single photons from such frequency-entangled photon pair source using a commercially available single-photon detector with a typical  time resolution of about 1 ns. Even the state-of-art single-photon detector with an ultra-high time resolution of about 20 ps is not fast enough. Therefore, disentangling the photon pairs into a factorable joint spectrum by shaping the pump field temporal modes has been considered as the only achievable method to produce heralded pure single-photon states from these wide-band sources without spectral filtering. This situation has been changed recently with the development of narrowband biphotons generation from SFWM in cold atoms \cite{DuJOSAB2008, HarrisPRL2005, DuPRL2008, DuOptica2014, KurtsieferPRL2013}  and cavity-enhanced SPDC \cite{PanPRL2008, ChuuAPL2012}. These biphotons, having bandwidths ranging from below 100 MHz down to sub-MHz and coherence time from 10 ns up to more than 1 $\mu$s, are ideal for generating heralded pure single-photon states using commercial detectors.

As the time origin is setted by the detection of the trigger photon ($t_2=0$), the heralded photon has a well-defined temporal wave function. To illustrate this, we work at $BW \Delta t \leq 1$ and thus it can be treated as an ideal heralding process. From Eq. (\ref{eq:HeraldedState0}), we obtain the temporal wave function of the heralded single photon:
\begin{eqnarray}
\label{eq:HeraldedWavefunction}
\psi_{1|2}(\tau)&=&\langle 0|\hat{a}_1(\tau)|\Psi_{1}\rangle_2|_{t_2=0} \nonumber \\
&=&\frac{1}{2\pi\sqrt{R}}\int d\Omega \Phi(\Omega)e^{-i(\omega_{10}+\Omega)\tau}   \nonumber \\
&=&\psi_0(\tau) e^{-i\omega_{10}\tau},
\end{eqnarray}
where $\psi_0(\tau)=\frac{1}{2\pi\sqrt{R}}\int d\Omega \Phi(\Omega)e^{-i\Omega\tau} $ is the Fourier transform of the two-photon joint spectrum function. Equation (\ref{eq:HeraldedWavefunction}) also clearly demonstrates that the quantum state of the heralded single photon is a coherent superposition of the frequency components in its spectrum, but not a mixed state as commonly believed. The time origin setted by the trigger photon allows for shaping the heralded single photons with an electro-optic modulator \cite{SinglePhotonEOM}. The recent experiments of single photons coherently interacting with atoms \cite{DuPRL2011, DuOE2012, DuPRL2012} and cavities \cite{DuPRL2014, KurtsieferPRL2014} have strongly evidenced that these heralded narrowband single photons indeed have coherent wave packets, but their quantum-state purities have never been formally and theoritically justified.

Following Eqs. (\ref{eq:HeraldedState0}) and (\ref{eq:HeraldedWavefunction}), we can produce many interesting heralded single-photon states by engineering the biphoton joint spectrum and frequency entanglement. For example, from a frequency-bin entangled two-photon state $|\omega_{10}+\delta\rangle|\omega_{20}\rangle+e^{i\theta}|\omega_{10}\rangle|\omega_{20}+\delta\rangle$, one can generate a heralded single-photon two-color qubit state $|\omega_{20}\rangle+e^{i\theta}|\omega_{20}+\delta\rangle$. This will certainly find important applications in quantum information processing and quantum communication.

\begin{figure}
\includegraphics[width=\linewidth]{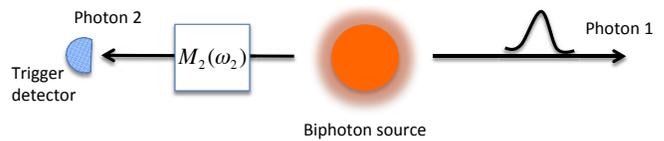}
\caption{(color online). Nonlocal modulation. A spectrum modulator $M_2(\omega_2)$ is placed on the path of the trigger photon 2 to nonlocaly shape the heralded photon 1.
\label{fig:NonlocalModulation}}
\end{figure}

As compared to the heralded photons generated from an frequency-uncorrelated source, heralding single photons from the frequency-entangled (anti-correlated) biphotons has two unique features and advantages. The first is that we can reverse the relative time by switching the trigger photon. In the above discussion, we take photon 2 as the trigger photon and obtain the heralded photon 1 with a temporal waveform $\psi_{1|2}(\tau)=\psi_0(\tau) e^{-i\omega_{10}\tau}$ given in Eq. (\ref{eq:HeraldedWavefunction}). Now let's switch the trigger detection to photon 1 to heralding the presence of photon 2. We obtain the temporal wave function of the heralded photon 2 as  $\psi_{2|1}(\tau)=\psi_0(-\tau) e^{-i\omega_{20}\tau}$, whose amplitude envelope is the time-reversal of the heralded photon 1. This time reversal feature, resulting directly from the frequency anti-correlation, have an important application. For example, when generated from an atomic multi-level system, the biphoton correlation can exhibit exponential decay waveforms because of the nature of spontaneous emission \cite{KurtsieferPRL2013}.  Depending on which photon is detected as the trigger,  we can generated heralded photons with an exponential growth or decay waveform that are particularly useful for loaded them into a cavity or interacting them with atoms \cite{DuPRL2012, DuPRL2014}. The second interesting feature is the nonlocal spectrum modulation \cite{BelliniPRA692004, HarrisPRA2008, HumblePRA2010}, which allows for shaping the heralded photon by placing a spectrum modulator $\textrm{M}_2(\omega_{2})$ on the trigger photon path, as illustrated in Fig. \ref{fig:NonlocalModulation}.  It can be derived that, the temporal wave function of the heralded photon becomes
\begin{eqnarray}
\label{eq:HeraldedWavefunctionM}
\psi_{1|2}(\tau)=\frac{1}{2\pi\sqrt{R}}\int d\Omega \Phi(\Omega)\textrm{M}_2(\omega_{20}-\Omega)e^{-i(\omega_{10}+\Omega)\tau}.  \nonumber \\
\end{eqnarray}
If the photon 1 experiences any distortion caused by the propagation dispersion, it can also be nonlocaly corrected by placing a dispersion compensation element on the path of the trigger photon 2.

In summary, we demonstrate that the detection of one photon from a frequency-entangled (frequency-anti-correlated) two-photon state, as the time origin established by the trigger photon detection has uncertainty much shorter than the two-photon coherence time, projects the remaining photon onto a pure single-photon state. The physics behind such a heralding process can be pictured as the following: the trigger detection erases the frequency information of the trigger photon, and as a result the heralded photon is in a superposition state of its frequency components. For the first time, we provide a full theoretical justification on the quantum-state purity of such heralded photons, while they had been commonly believed as mixtures of multiple frequency components. We show that the quantum-state purity depends on the response time of the trigger-photon detector. A purity of higher than 0.98 can be achieved as the trigger-detection response time is shorter than one-tenth of the biphoton coherence time (inverse of the bandwidth). The purity becomes significantly degraded as the trigger-detection response time is longer than the biphoton coherence time.  We also show that such a heralded narrowband single photon has a well-defined coherent wave packet. The entanglement raised from the frequency anti-correlation allows for reversing the relative time by switching the trigger photon and shaping the heralded single photon through nonlocal spectral modulation.

The work was support by the Hong Kong Research Grants Council (Project No. 16301214).


\end{document}